\documentclass[12pt,preprint]{aastex}

\shorttitle{SFHs and Ages of Massive ETGs at z $\sim$ 1.3}
\shortauthors{Rettura et al.}

\begin{document}


\title{Early-Type galaxies at z $\sim$ 1.3. III. On the dependence of Formation Epochs and Star Formation Histories on Stellar Mass and Environment}


\author{A. Rettura\altaffilmark{1,4,5}, S. Mei\altaffilmark{2,3},  S.A. Stanford\altaffilmark{1}, A. Raichoor\altaffilmark{2,3}, S. Moran\altaffilmark{4}, B. Holden\altaffilmark{6}, P. Rosati\altaffilmark{7}, R. Ellis\altaffilmark{8}, F. Nakata\altaffilmark{9}, M. Nonino\altaffilmark{10}, T. Treu\altaffilmark{11}, J.P. Blakeslee\altaffilmark{12}, R. Demarco\altaffilmark{13}, P. Eisenhardt\altaffilmark{14}, H.C. Ford\altaffilmark{4}, R.A.E. Fosbury\altaffilmark{7}, G. Illingworth\altaffilmark{6}, M. Huertas-Company\altaffilmark{2,3}, M.J. Jee\altaffilmark{1}, T. Kodama\altaffilmark{15}, M. Postman\altaffilmark{16}, M. Tanaka\altaffilmark{17}, R.L. White\altaffilmark{16}}

\altaffiltext{1}{Department of Physics, University of California, Davis, CA 95616, USA}
\altaffiltext{2}{University of Paris Denis Diderot, 75205 Paris Cedex 13, France}
\altaffiltext{3}{GEPI, Observatoire de Paris, Section de Meudon, Meudon Cedex, France}
\altaffiltext{4}{Department of Physics and Astronomy, Johns Hopkins University, Baltimore, MD 21218, USA}
\altaffiltext{5}{Department of Physics and Astronomy, University of California, Riverside, CA 92521, USA}
\altaffiltext{6}{UCO/Lick Observatories, University of California, Santa Cruz, CA 92065, USA}
\altaffiltext{7}{European Southern Observatory, 85748, Garching, Germany}
\altaffiltext{8}{California Institute of Technology, Pasadena, CA 91125, USA}
\altaffiltext{9}{Subaru Telescope, National Astronomical Observatory of Japan, Hilo, HI 96720, USA}
\altaffiltext{10}{INAF-Osservatorio Astronomico di Trieste, 34131 Trieste, Italy}
\altaffiltext{11}{Department of Physics, University of California, Santa Barbara, CA 93106, USA}
\altaffiltext{12}{Herzberg Institute of Astrophysics, National Research Council of Canada, Victoria,BC V9E 2E7, Canada}
\altaffiltext{13}{Department of Astronomy, Universidad de Concepcion, Casilla 160-C, Concepcion, Chile}
\altaffiltext{14}{Jet Propulsion Laboratory, California Institute of Technology, MS 169-327, Pasadena, CA 91109, USA }
\altaffiltext{15}{National Astronomical Observatory of Japan, Mitaka, Tokyo 181-8588, Japan}
\altaffiltext{16}{Space Telescope Science Institute, Baltimore, MD 21218, USA}
\altaffiltext{17}{Institute for the Physics and Mathematics of the Universe, The University of Tokyo, 5-1-5 Kashiwanoha, Kashiwa-shi, Chiba 277-8583, Japan}





\begin{abstract}
  We  study   the  environmental  dependence   of  stellar  population
  properties at  $z \sim 1.3$.   We derive galaxy  properties (stellar
  masses, ages  and star formation histories) for  samples of massive,
  red,  passive  early-type galaxies  in  two high-redshift  clusters,
  RXJ0849+4452 and RXJ0848+4453 (with redshifts  of z = 1.26 and 1.27,
  respectively),  and  compare  them   with  those  measured  for  the
  RDCS1252.9-2927  cluster at  z=1.24 and  with those  measured  for a
  similarly  mass-selected sample of  field contemporaries  drawn from
  the  GOODS-South  Field.   Robust  estimates of  the  aforementioned
  parameters have been obtained by comparing a large grid of composite
  stellar  population  models  with  extensive 8-10  band  photometric
  coverage, from  the rest-frame far-ultraviolet to  the infrared.  We
  find  no variations  of  the overall  stellar population  properties
  among the different samples of cluster early-type galaxies. However,
  when comparing cluster versus field stellar population properties we
  find that,  even if the  (star formation weighted) ages  are similar
  and depend  only on  galaxy mass,  the ones in  the field  do employ
  longer  timescales to  assemble  their final  mass.   We find  that,
  approximately 1 Gyr after the  onset of star formation, the majority
  (75\%) of cluster galaxies have already assembled most ($> 80\%$) of
  their final mass,  while, by the same time,  fewer (35\%) field ETGs
  have.  Thus we conclude that  while galaxy mass regulates the timing
  of  galaxy formation,  the  environment regulates  the timescale  of
  their star formation histories.

\end{abstract}





\section{Introduction}

Galaxies reside in environments that  span a wide range of density. In
order to understand the  physical processes that drive their evolution
is important  to test for  systematic differences between  galaxies in
various environments.  Many authors have shown that population density
plays an important role in determining many galaxy properties, such as
star     formation     rate,     gas    content     and     morphology
\citep{Kodama01,Balogh02}.  Several  mechanisms have been  proposed by
theorists  to  account  for   these  effects,  such  as  ram  pressure
stripping,  mergers  and   tidal  effects  \citep{Gunn72,  Dressler97,
  Moore96,Moore98,Moore99}.

More than half of all stars in the local Universe are found in massive
spheroids (e.g.,  \citet{Bell03}).  The  history of mass  assembly and
star formation of these Early-Type  Galaxies (ETGs) are among the most
actively pursued elements in galaxy evolution studies, and form the
basis  for  models  of  massive  galaxy  formation  \citep{Renzini06}.\\
Galaxy  clusters  provide  an  ideal  laboratory to  study  ETGs,  the
dominant galaxy population in  clusters, even beyond redshift one.  In
addition,  large  field  surveys  have  multiplied  in  recent  years,
extending  the  baseline  over  which  environmental  effects  can  be
studied .

Studies indicate that the most massive  ETGs {\it in the field} may be
amongst  the oldest  at any  given epoch  \citep{Cimatti04, Fontana04,
  Saracco04, Treu05,  Juneau05,dSA05, Pannella09}.  They  have evolved
mainly passively since $z \sim 1$ and, at least the most massive ones,
as slowly as  cluster galaxies \citep{vandokkum96, Bernardi98, Treu99,
  Kochanek00, vandokkum01,  Treu01, vandokkumstanford03, Depropris07}.
Similarly, tight constraints have been placed on the scatter and slope
of  the color-magnitude  relation  (CMR), the  Fundamental Plane  (FP)
\citep{Jorgensen06, Treu05, vdW05, dSA06, vandokkum07} and Balmer line
strengths  \citep{Clemens06, Sanchez06}  of the  ETGs  populating {\it
  massive clusters}, indicating a very high formation redshift for the
ETGs  in clusters as  well \citep{Rettura06,  Eisenhardt08, Rettura10,
  Gobat08, Mei09, Collins09, Rosati09}.

The most distant clusters known to date provide the strongest leverage
on  model  predictions  \citep{Papovich10, Tanaka10a}.   Specifically,
there is  a tight CMR  of ETGs at  $0.8< z <  1.5$ \citep{Blakeslee03,
  Blakeslee06,  Mei06a, Mei06b,  Lidman08, Hilton09,  Strazzullo10}, a
slowly evolving  $K$-band luminosity  function seemingly at  odds with
hierarchical  merging scenarios  \citep{Toft04,  Strazzullo06}, and  a
tight and slowly evolving FP out to $z\approx 1.2$ \citep{Hol05}.  The
existence  of such  massive,  passively evolving  galaxies already  at
$z\gtrsim 0.8$  needs to be  reconciled with the evidence  for massive
(often dusty) star-forming populations  found at $z\gtrsim2.2$ both in
the   field   and    in   overdense   environments   \citep{Steidel05,
  Adelberger05, Kodama07, Miley08, Overzier09, Tanaka10b}.

Previous studies based  on optical and infrared data  have placed some
constraints  on  the  star  formation  histories of  ETGs  in  distant
clusters based on the scatter in their CMRs: the best-fit models yield
formation epochs of $2 < z_{f} < 7$ \citep{Blakeslee03}, but a greater
accuracy  could not  be  achieved based  on  the available  rest-frame
optical  and  infrared  data  alone.   The reason  for  this  is  that
age-dependent features  in galaxy  spectra (e.g., the  4000 \AA\mbox{}
break or the 1.6  $\mu m$ bump) are too broad or  evolve too slowly to
be  used  as  adequate  age indicators  \citep{Burstein88,  Tantalo96,
  Maraston05, Rettura10}.  In order  to better constrain the formation
epoch  ($z_{f}$)  of  ETGs we  need  to  probe  the much  sharper  and
fast-evolving  rest frame  UV-optical  colors that  will  allow us  to
determine $z_{f}$  with much greater accuracy  than hitherto achieved,
enabling us  to better constrain the entire  SFH of ETGs\footnote{Note
  that by  $z\gtrsim 1.3$  not enough cosmic  time has elapsed  for HB
  stars to  produce the ``UV upturn'' \citep{Yi97,  Yi99}.}  (see also
Fig.2 of \citet{Rettura10})

In  order to  provide a  key test  of the  paradigm of  an accelerated
evolution  in  the  highest  density  environments  \citep{Diaferio01,
  Thomas05}, in this work, on  the basis of photometric data available
over the rest-frame wavelength  range $0.15-2\mu$m, we compare stellar
masses, ages  and inferred star  formation histories of ETGs  found in
three massive  X-ray detected clusters at $z  \sim 1.3$ (RXJ0849+4452,
RXJ0848+4453, RDCS1252.9-2927) with a sample of field contemporaries
drawn from  the GOODS South Field.

The structure  of this  paper is as  follows.  The description  of our
datasets, cataloging and sample selection is described in \S 2. In \S
3 we describe our methods  in inferring star formation histories, ages
and masses from stellar population analysis.  The results of our study
are discussed in \S 4, while  in \S \hspace{0.05cm} 5 we summarize our
conclusions.

We assume a $\Omega_{\Lambda} = 0.73$, $\Omega_{m} = 0.27$ and $H_{0}
= 71\ \rm{km} \cdot \rm{s}^{-1} \cdot \rm{Mpc}^{-1}$ flat universe
\citep{Spergel03}, and use magnitudes in the AB system throughout this
work.

\section{Description  of  the  data}

This work  is based  on data collection  programs on the  three galaxy
clusters  at  $z  \sim  1.3$  which are  amongst  the  most  extensive
spectroscopic  and photometric surveys  available over  the wavelength
range $0.35-4.5\mu$m.   To compare cluster galaxy  properties with the
field,  we  also  take  advantage  of  similar-quality  archival  data
available  for the  GOODS-South Field.   Tab.~\ref{table:1} summarizes
the rich and homogeneous dataset employed in this study.


\subsection{RX J0849+4452 and RX J0848+4453}
The  Lynx Supercluster,  is  the highest  redshift supercluster  known
today  \citep{Nakata05},  with   two  central  X-ray  detected  galaxy
clusters  and three  spectroscopically  confirmed surrounding  groups.
This work focuses in particular on the cores of the two main clusters,
RX J0849+4452  (hereafter, Lynx E) and RX  J0848+4453 (hereafter, Lynx
W), both  detected in the  ROSAT Deep Cluster  Survey \citep{Rosati99}
and spectroscopically confirmed at z  = 1.261 \citep{Rosati99} and z =
1.273   \citep{Stanford97},   respectively.    Furthermore,   in   two
accompanying  papers \citep{Raichoor11,  Mei11} we  also  compare ETGs
stellar population properties  in the two cluster cores  with those of
the ones in the surrounding groups.

Both Lynx E  and Lynx W were observed in  the $u^{\prime}$ filter with
the blue  channel of LRIS on Keck  I in two separate  epochs. On March
4th, 2003, exposures totaling 260 min were obtained in conditions that
were photometric with  $0.8"$ seeing.  A further 240  min was obtained
during the  night of March 1st,  2008, with clear skies  and a typical
seeing  of $0.9"$.   The two  epochs of  data were  reduced separately
using  standard  IRAF  tasks,  where  we subtracted  the  bias  level,
flat-fielded each  exposure using a series of  twilight flats obtained
on each  night, astrometrically calibrated  and aligned each  image by
comparison  to star  positions from  the HST  GSC2.2 catalog  and then
refined with  a larger number of  stellar positions from  the SDSS DR6
catalog. Finally, the images were flux-calibrated, also correcting for
atmospheric  extinction,   using  exposures  of   various  photometric
standard stars  obtained across a  range of airmasses each  night, and
then  combined.   The two  sets  of  images  were taken  at  different
position angles, and so the combined frame has an effective exposure time
of 8 hrs 20 min  across much of the field, but a  shallower depth at the
edges of the field where the frames  do not overlap.   The effective
resolution in the combined ultraviolet image is $0.95"$.

In order to sample  the optical-to-infrared wavelength domain, we have
obtained   data   in   the   following  passbands,   R,   $i_{F775W}$,
$z_{F850LP}$, J, Ks, and  $Spitzer$/IRAC channels, 3.6$\mu$m (ch1) and
4.5$\mu$m (ch2).   \\ We  refer the reader  to the companion  paper by
\citep{Raichoor11} for a more  detailed description of the optical and
infrared dataset used in this work (see also their Tab.1).

Here we remind that the R-band images were obtained with the Keck/LRIS
instrument.   The $i_{F775W}$ and  $z_{F850LP}$-band were  observed in
the $F775W$ and $F850LP$ filters provided by the ACS-Wide Field Camera
(WFC) on board of the Hubble  Space Telescope.  We note that these two
observing filters have been purposely chosen to bracket the
4000$\AA$-break of a model elliptical galaxy at $z = 1.26-1.27$ \citep{Mei06a, Mei06b}.\\
The near-infrared imaging data ($J, K_s$-band) were acquired with the
FLAMINGOS instrument available at the Kitt Peak National Observatory (KPNO).\\
Mid-Infrared imaging was obtained  (in two channels at [3.6$\mu$m] and
[4.5$\mu$m])  with the  IRAC camera  on board  of the  $Spitzer$ space
telescope.

\subsection{RDCS1252.9-2927 and the GOODS-South field.}

This work builds  on datasets and analyses already  performed on ETGs
belonging to  the X-ray luminous  cluster, RDCS1252.9-2927 (hereafter,
CL1252),  and to  the  GOODS-South field,  centered  on the  so-called
Chandra  Deep Field  South  field (hereafter,  CDFS),  as reported  by
\citet{Rettura10, Gobat08}.

We  refer to the  aforementioned papers  for more  details in  the data
reduction. Here we note that the  data we have employed for the CL1252
cluster consist of deep imaging in 10 bands: \textit{VLT}/VIMOS ($U$),
\textit{VLT}/FORS2 ($B$,  $V$, $R$), \textit{HST}/ACS
($i_{F775W}$,$z_{F850lp}$)   \citep{Blakeslee03},   \textit{VLT}/ISAAC
($J_s$,$K_s$)     \citep{Lidman04},     \textit{Spitzer}/IRAC    ($3.6
\mu\mbox{}m$, $4.5 \mu\mbox{}m$), as  well as spectroscopic data taken
with \textit{VLT}/FORS2 and published in
\citet{Demarco07}.

The archival data  for the comparison field, the  CDFS, comprises deep
imaging  in   9  bands:  \textit{VLT}/VIMOS   ($U$)  \citep{Nonino09},
\textit{HST}/ACS ($B_{F435W}$,  $V_{F606W}$, $i_{775W}$, $z_{F850LP}$)
\citep{Giava04},  \textit{VLT}/ISAAC,  ($J$,$K_s$) \citep{Retzlaff10},
\textit{Spitzer}/IRAC    ($3.6   \mu\mbox{}m$,    $4.5   \mu\mbox{}m$)
\footnote{CDFS  imaging  is   publicly  available  through  the  GOODS
  collaboration  web-site:  http://www.stsci.edu/science/goods/.},  as
well as spectroscopic data taken with the \textit{VLT}/FORS2 300I grism by
the   ESO-GOODS   survey\footnote{Spectroscopic   data  are   publicly
  available              through             the             web-site:
  http://www.eso.org/sci/activities/projects/goods/MasterSpectroscopy.htm.}
\citep{Vanzella05,Vanzella06,Vanzella08,Balestra10} and the K20 survey
\citep{Cimatti02}.

\subsection{Cataloging of  observations  and samples selection}

The  resulting  datasets  for the three  clusters  and  the  field  have
homogeneous depths  and wavelength coverage,  allowing the application
of similar  selection criteria  for both samples.   This is  a crucial
point  for stellar population  studies, so  that we do  not have
different levels of systematic biases in the analyses.

The   data  allow   the  reconstruction   of  galaxy   spectral  energy
distributions (SEDs)  by entirely sampling  the relevant  spectrum range
emitted by all the  different stellar populations.  For each observing
band we  produce catalogs from  matched aperture photometry,  with an
aperture of radius 1.5'' and an aperture correction out to 7'' radius,
as described in \citet{Rettura06, Raichoor11}.

The availability of 8 to 10 passbands spanning such a large wavelength
range    enables   the   estimate    of   accurate    stellar   masses
\citep{Rettura06},   ages  and  star   formation  histories   of  ETGs
\citep{Rettura10} and enables us to directly compare galaxy properties
of homogeneously selected samples of ETGs in both environments.

Specifically, we  measure stellar population parameters of  Lynx E and
Lynx W early-type  cluster galaxies and compare them  with those found
for  similarly-selected samples  of cluster  and  field contemporaries
drawn from CL1252 at $z= 1.237$ and CDFS at $z= 1.237 \pm 0.15$.

We estimated  stellar masses from  both $K_{s}$ band  observations and
SED fitting .  The two  estimates give similar mass limits.  The depth
of the  {\it VLT}/ISAAC  and {\it KPNO}/FLAMINGOS  $K_{s}$-band images
and the  extended multi-wavelength data  for all fields allows  us, in
fact,  to  define  complete  mass-selected  samples.   The  reader  is
referred  to \citet{Gobat08,Raichoor11}  for more  details.   Here, we
note that overall photometric completeness is obtained if we limit our
analysis  to stellar  masses larger  than $M_{lim}  = 5  \cdot 10^{10}
M_{\odot}$\footnote{Assuming  Salpeter  Initial  Mass Function  (IMF),
  with lower  and upper cutoffs, $m_{L}=0.1  M_{\odot}$ and $m_{U}=100
  M_{\odot}$,    respectively.     We     refer    the    reader    to
  \citet{Rettura06,Raichoor11} for  a discussion on  the dependence of
  inferred stellar  population properties on the actual  choice of the
  IMF.}.

A  selection of  Lynx E,  Lynx  W and  CL1252 passive  ETGs along  the
cluster red sequence  is efficiently provided by a  color selection of
$i_{F775W}  -   z_{F850LP}>0.8$  \citep{Blakeslee03,  Mei06a,  Mei06b,
  Mei09,  Mei11}.\\  For  CL1252,   in  the  spectroscopic  sample  of
\citet{Demarco07},  there are 22  red-sequence galaxies  ($i_{F775W} -
z_{F850LP}>0.8$) with $M_{*} > M_{lim}$, of which 18 are classified as
passive ETGs, i.e. without detectable [OII] or other emission lines in
their  observed spectra.   For  Lynx E  and  Lynx W,  in the  combined
photometric  and   spectroscopic  samples  \citep{Mei06a,  Stanford97,
  Rosati99,  Stanford01, vandokkumstanford03,  Holden11} there  are 21
red-sequence  galaxies  with  $M_{*}  >  M_{lim}$,  of  which  17  are
classified  as  passive  ETGs.   Specifically, there  are  10  passive
spectroscopically-confirmed  ETGs in  Lynx E,  and  3 in  Lynx W.   We
remark that, to improve the statistics, we have also included into the
Lynx ETGs sample, 3 photometrically-confirmed  members of Lynx E and 1
photometrically-confirmed  member of  Lynx  W both  with $i_{F775W}  -
z_{F850LP}>0.8$ and $M_{*}>M_{lim}$, thus  resulting in a total sample
of 17  Lynx E+W ETGs.  It is  important to note that  the removal from
our sample  of these 4 galaxies, whose  cluster-membership was assessed
on  the basis  of a  (8-9 band)  photometric-redshift only,  would not
change  the main conclusions  of this  study.\\ For  the corresponding
CDFS  sample  of field  contemporaries,  the  same  criteria yield  27
passive ETGs in  CDFS with FORS2 spectra giving  redshift in the range
$z  =  1.237  \pm  0.15$  (for more  details  see  also  \citet{Gobat08,
  Rettura10}).

We also remark that, adopting  the classification scheme of Postman et
al.   (2005),  visual morphological  analysis  of  the HST/ACS  images
available for  all samples  indicates that all  our red,  passive ETGs
also show typical elliptical or lenticular morphology.

As  reported  in  Rettura  et  al.   2010, we  also  remind  that  the
spectroscopic follow-up  for CL1252 is  more complete at  the low-mass
end than in CDFS. Thus, our sample of ETGs in CDFS is likely to be more
incomplete at the  low-mass end than the CL1252 and  Lynx ones. We will
return to this point when discussing our results in Section 4.

\section{Data Analysis: derivation of stellar population properties}

The  data  described  in  the  previous  section  are  used  to  infer
fundamental physical properties  of similarly mass-selected samples of
early-type galaxies (13 ETGs in the Lynx  E, 4 in Lynx W, 18 in CL1252
and   27  in   the   CDFS)   .   Adopting   a   similar  approach   to
\citet{Rettura06, Rettura10}, we derive stellar masses, star formation
histories  and  ages  for  each  ETG  in  Lynx  E  and  Lynx  W  using
multi-wavelength  PSF-matched aperture  photometry  from 8  passbands,
from observed $u'$-band to observed  $4.5 \mu m$.  For each galaxy, we
compare the observed  SED with a set of  composite stellar populations
(hereafter,  CSP)  templates computed  with  the \citet{BC03}  models,
assuming solar  metallicity, \citet{Salpeter55} Initial  Mass Function
(hereafter, IMF) and no dust.

Similarly  to   \citet{Rettura06}  we  checked  the   effect  of  dust
extinction on the  best-fit stellar masses by including  a fourth free
parameter,  $0.0<  E(B-V)<   0.4$,  following  the  \citet{Cardelli89}
prescription.  Performing  the SED fit on  13 Lynx E+W  ETGs (all with
complete $u'$-to-$4.5\mu  m$ coverage) and on  22 CDFS ETGs (all with
complete $U$-to-$4.5\mu  m$ coverage),  we find  that in  $ \sim
50\%$ of  the cases  $E(B-V) \leq  0.05$ gives the  best fit.   In the
remaining  cases  values  of  $E(B-V)  \leq  0.2$  are  found  and  no
particular  trend with  mass nor environment is found  either,  hence supporting  the
validity of the dust-free assumption we make throughout this work.


For  our CSP  models, we assume  the following  grid of
exponentially-declining star formation  history (SFH) scenarios, $\Psi
(t, \tau)$:


\begin{equation} \label{eq:sfhs}
\Psi (t, \tau) = SFR_{0} \cdot e^{-t/\tau} \hspace{0.5cm} \Bigg[ \frac{M_{\odot}}{yr} \Bigg],
\end{equation}


where $0.05 \leq \tau \leq 5$ Gyr, $SFR_{0}$ is the initial SFR at the
onset of star  formation, and $t$ is the time since  the onset of star
formation\footnote{The range of acceptable $t$s for a given galaxy has
  been limited by  the age of the universe  at its observed redshift.}
of the stellar population model formed at a lookback time $T(\overline
z)+t$, with,  e.g., $T  \simeq 8.67$ Gyr  at the epoch  of observation
($\overline z \simeq 1.3$).

In determining galaxy model  ages, masses and star formation histories
from SED  fitting, is important  to understand how much  our estimates
could possibly  be affected by  ``age-metallicity''\footnote{We employ
  the  working assumption that  the most-massive  ETGs have  all solar
  metallicities.}   and  ``age-SFH'' degeneracies.   We  note that  in
\citet{Rettura10}  we have  demonstrated that  the use  of photometric
information coming  from the rest-frame  UV is crucial  to distinguish
the different parameters of  the stellar population modeling, allowing
us to break the ``age-SFH  degeneracy''.  It is also important to note
that  the  rest-frame  UV  remains  also  largely  unaffected  by  the
``age-metallicity''   degeneracy,   which   plagues  optical   studies
\citep{Worthey94}.

Furthermore,  the  reliability, at  $z  >  1$,  of stellar  population
parameters inferred in the  rest-frame NIR regime ($\lambda_{obs} \sim
2\mu  m$) has  long  been debated  \citep{Maraston98}  because of  the
different  implementation  of  the relevant  short-duration  thermally
pulsating (TP) AGB phase in the different stellar population synthesis
models.  However,  in \citet{Rettura06}  we have shown  that different
stellar  population models actually  yield overall  consistent stellar
masses, within  the typical errors, for  ETGs at $z \sim  1.0$.  It is
clear  that the  inferred  SFH  (and age)  is  model-dependent and  is
generally not  a unique solution,  however the relative  difference in
the underlying  stellar population is  still significant. In  fact, in
\citet{Rettura10} we  have shown that when comparing  {\it cluster vs.
  field}  $\tau$s  and  ages,  the {\it  relative}  differences  among
samples (or lack thereof, as in their Fig.  8) were similar regardless
of  the  actual stellar  population  code  used,  not affecting  their
conclusions.   \\   Hence,  since  our   study  is  indeed   aimed  at
constraining  the  relative difference  of  galaxy stellar  population
parameters  in a  {\it  Cluster  vs.  Cluster}  and  {\it Cluster  vs.
  Field} fashion, we will only show here the results obtained with the
\citet{BC03}  models, in  order to  facilitate direct  comparison with
previous studies  in the literature.  Note  that in \citet{Raichoor11}
we also present  an extensive discussion on systematics  due to choice
of the stellar population models at $z \sim 1.3$ .

To account for the  average age of the bulk of the  stars in a galaxy,
we refer throughout this  paper to {\it star-formation weighted} ages,
$< t >_{SFH}$, defined as:


\begin{equation} \label{eq:age2}
< t >_{SFH} \equiv \frac{\int_{0}^{t} (t-t^{'}) \Psi(t^{'}, \tau) dt^{'}}{\int_{0}^{t} \Psi(t^{'}, \tau) dt^{'}} .
\end{equation}
Assuming $\Psi(t^{'}, \tau)$ as in Eq.~\ref{eq:sfhs} we obtain,
\begin{equation} \label{eq:age}
< t >_{SFH}  = \frac{t-\tau+\tau \cdot e^{-\frac{t}{\tau}}}{1-e^{-\frac{t}{\tau}}},
\end{equation}
where $t$ is the time elapsed  since the onset of star formation (SF).

By  comparing  each  observed  SED  with these  atlases  of  synthetic
spectra, we construct  a 3D $\chi^{2}$ space spanning  a wide range of
star formation  histories, model ages and stellar  masses.  The galaxy
mass in  stars $M_{*}$, the inferred  $< t >_{SFH}$ and  the $\tau$ of
the  models giving  the  lowest  $\chi^2$ are  taken  as the  best-fit
estimates  of  the  galaxy   stellar  mass,  age,  and  SFH  timescale
throughout this work.
We note that this procedure  results in typical errors for galaxy ages
of $\sim  0.5$ Gyr, and for $\tau$ of  $\sim 0.2$ Gyr \citep{Rettura10}.
Typical uncertainties on the  mass determination are about $\sim 40\%$
(i.e., $0.15$ dex) \citep{Rettura06}.

\section{Results and Discussion}

\begin{figure*}
\epsscale{1.0}
\includegraphics[scale=.45]{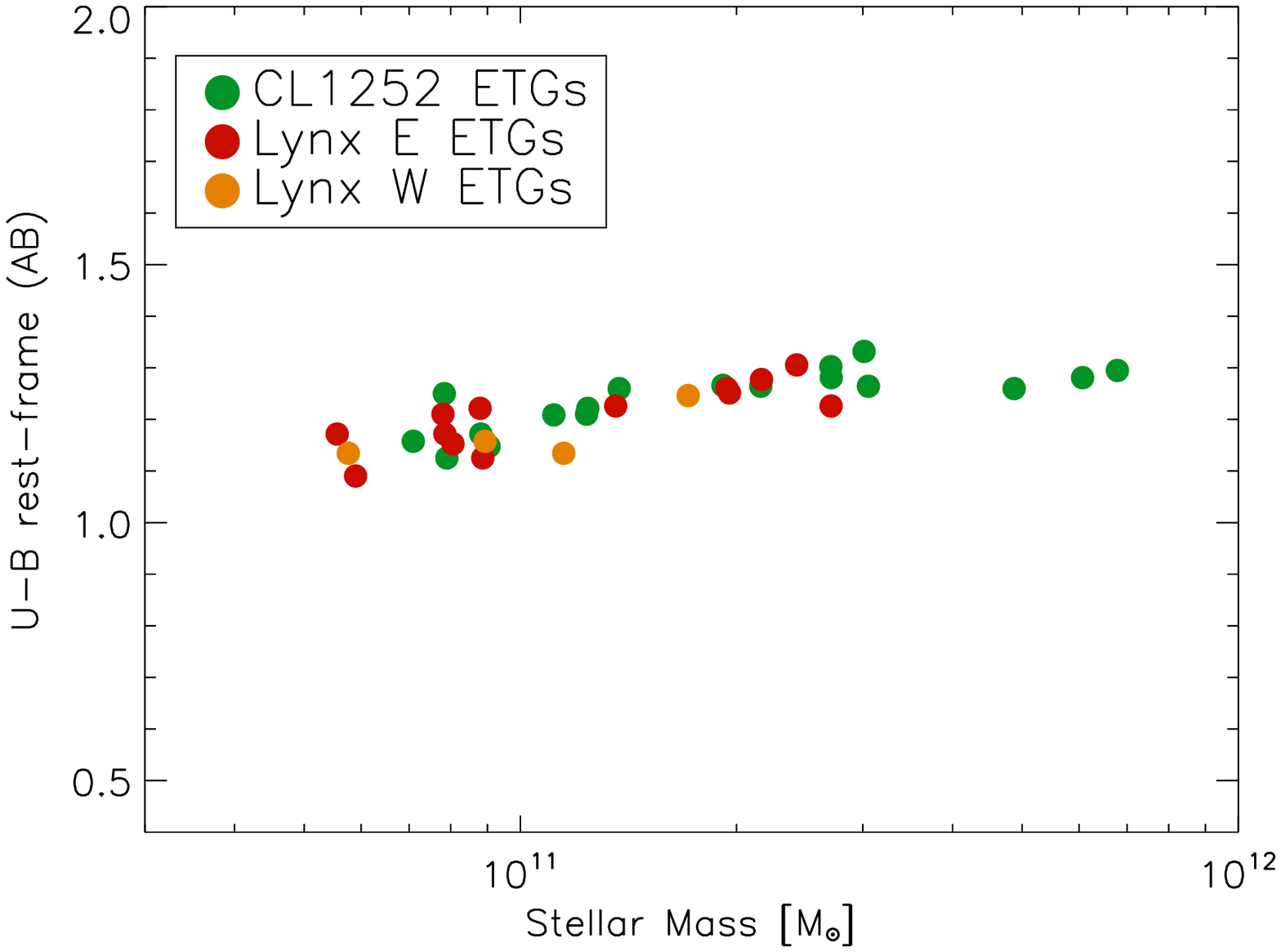}
\includegraphics[scale=.45]{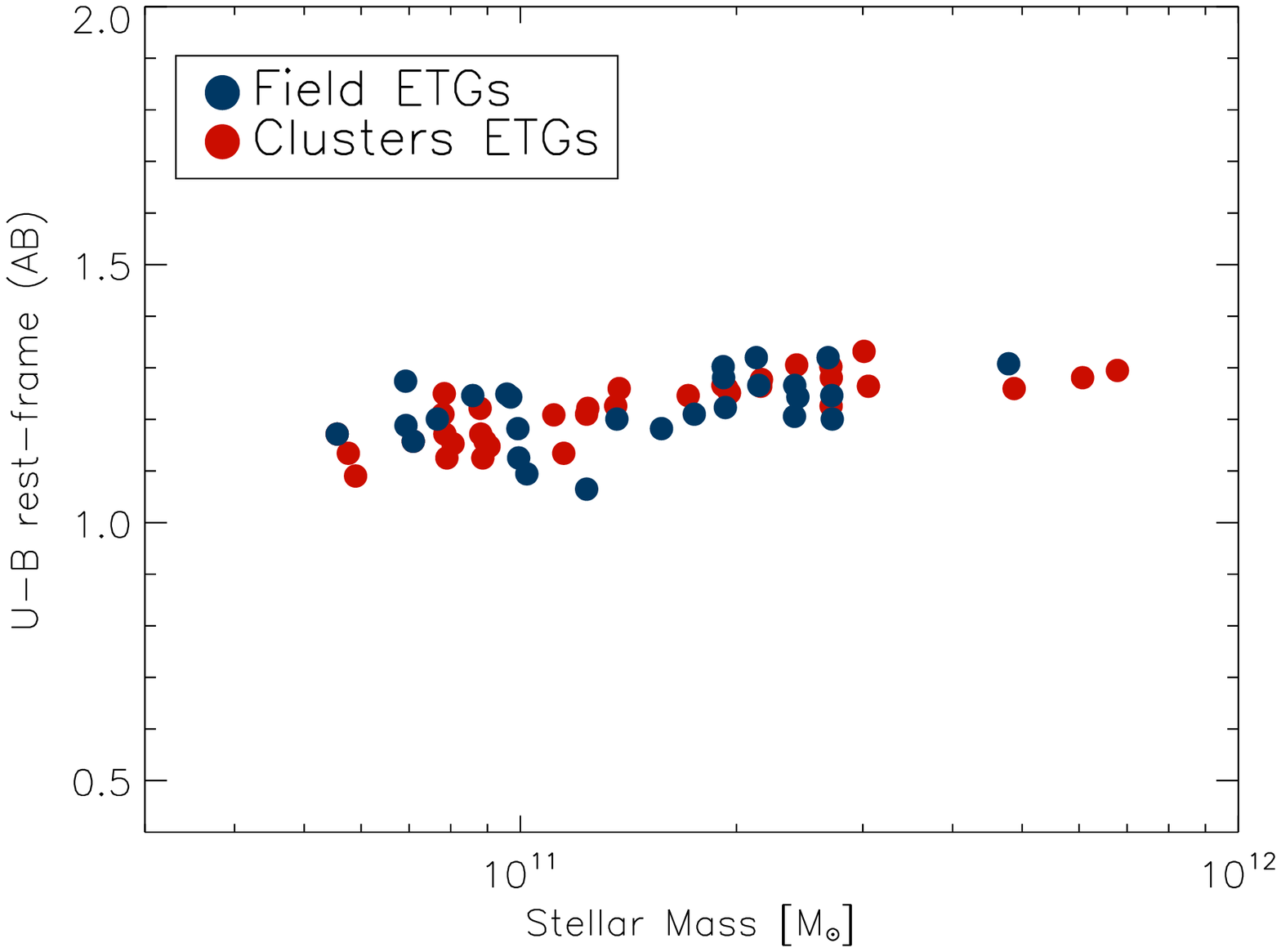}
\caption{{\it  Top  panel}:  Rest-frame  $U-B$ color-mass  diagram  of
  mass-selected  samples  of  CL1252  (green  circles),  Lynx  E  (red
  circles)  and   Lynx  W  (orange  circles)   passive  cluster  ETGs.
  Uncertainties in the  stellar mass are $\sim 0.15$  dex. {\it Bottom
    panel:}  $U-B$ color  - mass  diagram of  the combined  samples of
  Cluster  (red circles) and  Field (blue  circles) ETGs.   Field ETGs
  galaxies are  distributed around the  cluster red-sequence, although
  are found with a larger scatter.}
\label{UMB_mass}
\end{figure*}

In the top  panel of Fig.~\ref{UMB_mass} we plot  the inferred stellar
masses  as  a   function  of  the  $U-B$  rest-frame   color  for  the
mass-selected samples of CL1252, Lynx  E and Lynx W ETGs, resulting in
a   very    similar   distribution.    In   the    bottom   panel   of
Fig.~\ref{UMB_mass} we show the  rest-frame $U-B$ color - mass diagram
of  the combined  samples of  cluster  (red circles)  and field  (blue
circles) ETGs.  To  have a very simple estimate of  the scatter of the
color--mass relation, as  derived from the SED fitting,  we derive the
scatter around  the fit using a Tukey's  biweigth \citep{Press92}.  We
simply  calculated  the  uncertainty  in  the  scatter  estimation  by
bootstrapping  on 1,000  simulations.   For the  Lynx cluster  sample,
CL1252, and field galaxies, we  obtain a scatter of $0.031 \pm 0.010$,
$0.042  \pm 0.010$ ,and  $0.050 \pm  0.008$, respectively.   These are
intrinsic scatters as predicted from  the best SED fitting, and do not
take into  account uncertainties in  the fitting methods.   Field ETGs
galaxies  are distributed  around the  cluster  red-sequence, although
they seem  to show  a larger scatter,  especially at lower  masses. We
will  need more statistics  and better  estimates of  uncertainties to
draw  a stringent  conclusion, since  the cluster  and  field overall
scatters could still be consistent within the uncertainties.


As  we apply the  method described  in section  \S 3,  we are  able to
directly   compare  the  relative   distribution  of   galaxy  stellar
population properties for
the  different  samples.\\
Stellar masses  ranges from $M_{lim}$  to $4 \cdot  10^{11} M_{\odot}$
except for  three CL1252 bright galaxies  and one galaxy  in the field
(bottom panel of Fig.~\ref{UMB_mass}). We  do not find any of the Lynx
clusters ETGs to be more massive than $3 \cdot 10^{11} M_{\odot}$ (top
panel of Fig.~\ref{UMB_mass}).

We  want  to  tests  the  hypothesis  that  two  samples  stellar-mass
distributions have  the same median  against the hypothesis  that they
differ  at  the  5\%  significance  level.   The  Mann-Whitney  U-Test
suggests  that the  distributions  are likely  to  be similar  (median
probability of  44\%). We also study  the Kolmogorov-Smirnov statistic
and associated probability that the  two arrays of data are drawn from
the same distribution, yielding a  similar result. A high value (67\%)
of probability shows that  the cumulative distribution function of the
cluster stellar  masses is not significantly different  from the field
one.

\begin{figure*}
\epsscale{1.0}
\includegraphics[scale=.35,angle=90]{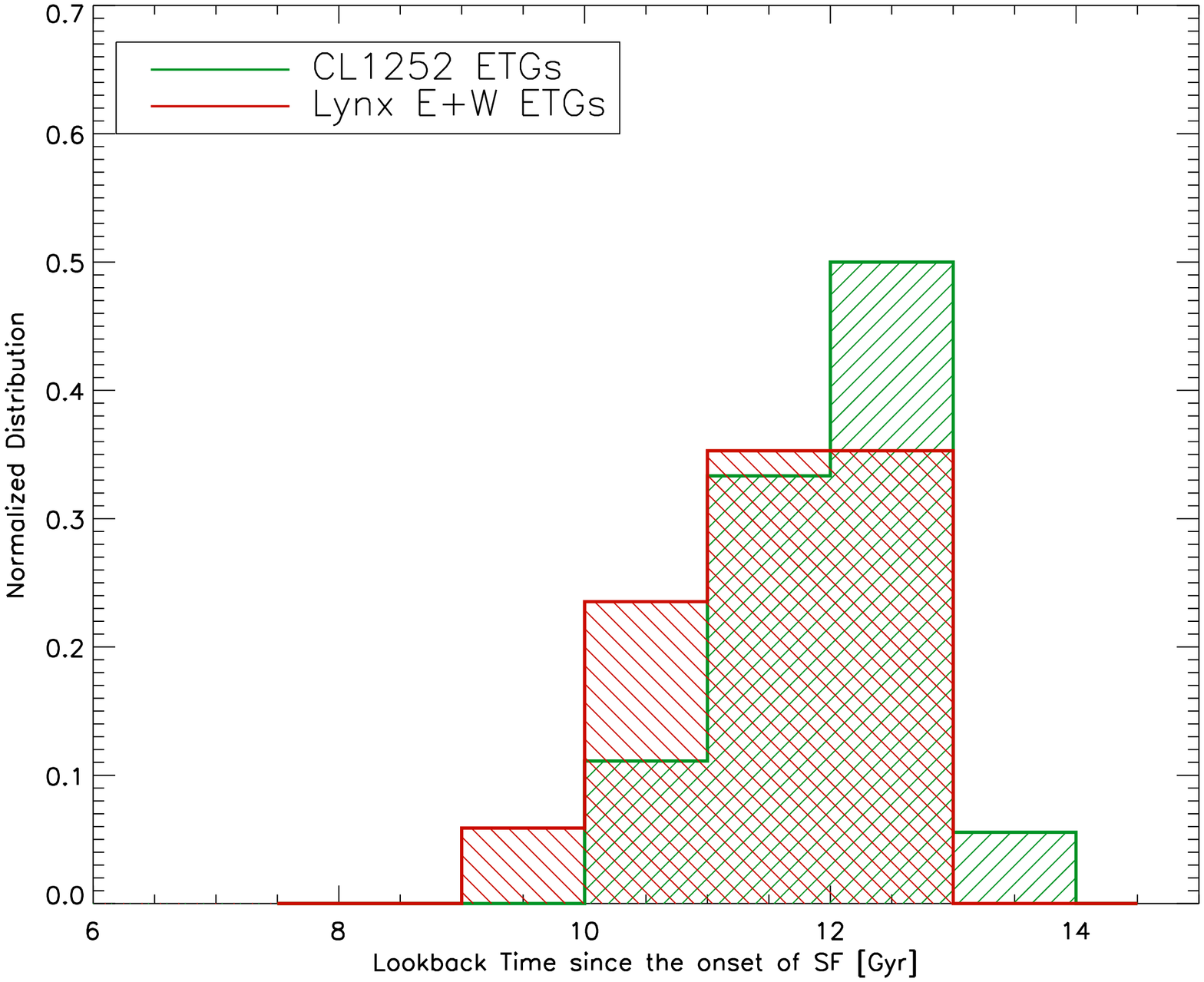}
\includegraphics[scale=.35,angle=90]{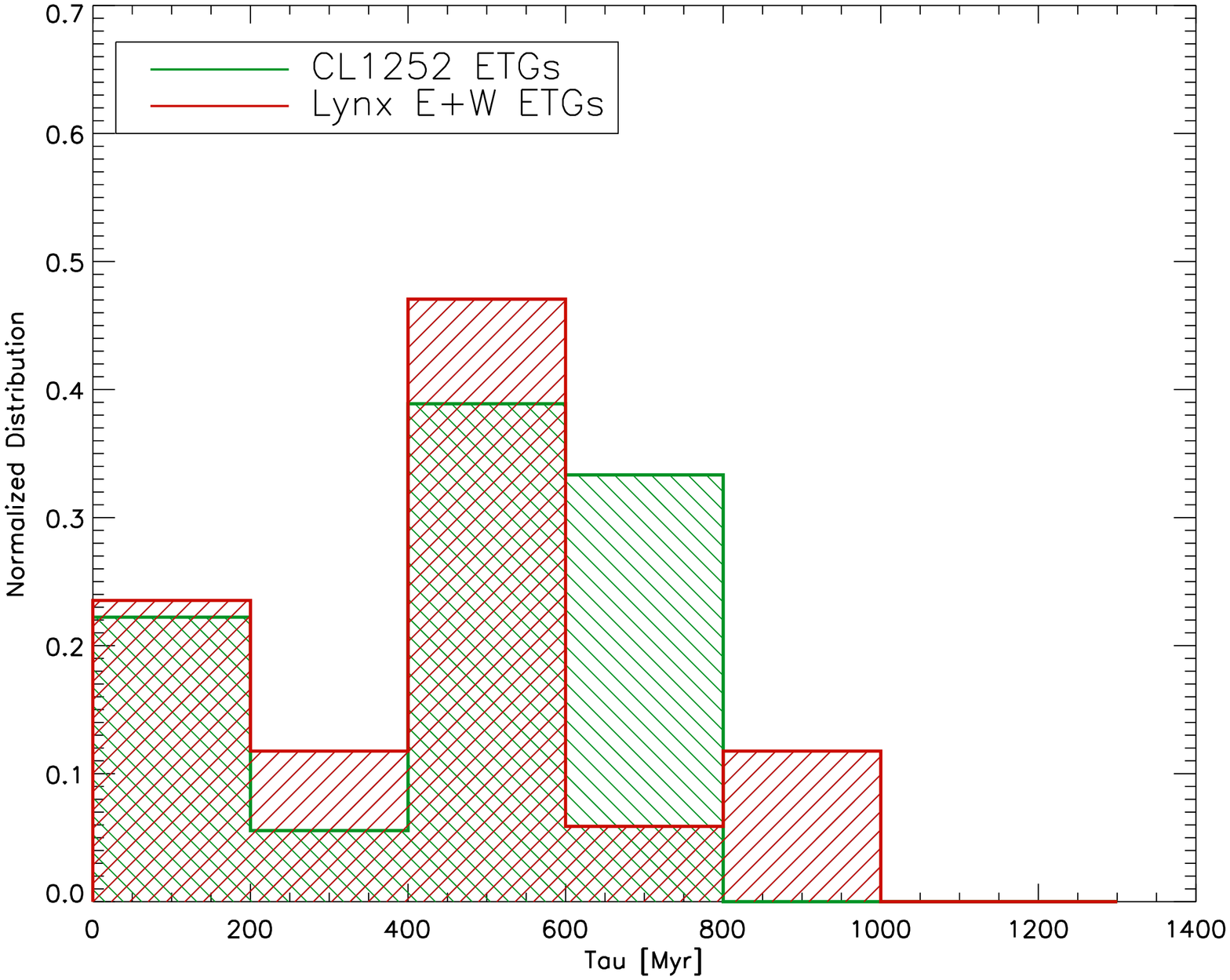}
\caption{{\it  Top Panel:}  Histograms of  the lookback  times  to the
  onsets of  star formation for the cluster  samples. Uncertainties in
  ages  are $\sim 0.5$  Gyr.  No  cluster-to-cluster variation  of the
  galaxy  ages   is  found.    {\it  Bottom  Panel:}   Formation  {\it
    timescales}, $\tau$, histograms for the different cluster samples.
  No cluster-to-cluster  variation of the  galaxy formation timescales
  is  found.  Uncertainties  in stellar  masses and  $\tau$  are $\sim
  0.15$ dex and $\sim 0.2$ Gyr, respectively.}
\label{look_and_tau_1252lynx}
\end{figure*}

As shown  in Fig.~\ref{look_and_tau_1252lynx},  we also find  that all
clusters show  the same relative distributions of  lookback time since
the  onset   of  star  formation  ($T(\overline   z)+t$,  top  panel)
and stellar mass assembly  timescales, ($\tau$; bottom  panel).  In fact  both the
Mann-Whitney U-Test (median  probability of   10\%, 21  \%) and the
K-S statistics (75\%,  91\% of probability) also elucidate that
the distributions are very similar.

This lack of cluster-to-cluster variation may well indicate that these
three clusters  experienced similar evolutionary  processes, resulting
in  similar  stellar  population  properties  shown  by  their  member
ETGs. Since these samples of cluster ETGs are found to be very similar
from a stellar population point-of-view,  we can combine them into one
sample of  35 ETGs,  representative of the  Cluster environment  at $z
\sim  1.3$,  apt for  comparison  with the  sample  of  27 CDFS  ETGs,
representative of the Field environment.

\begin{figure*}
\epsscale{1.}
\includegraphics[scale=.45]{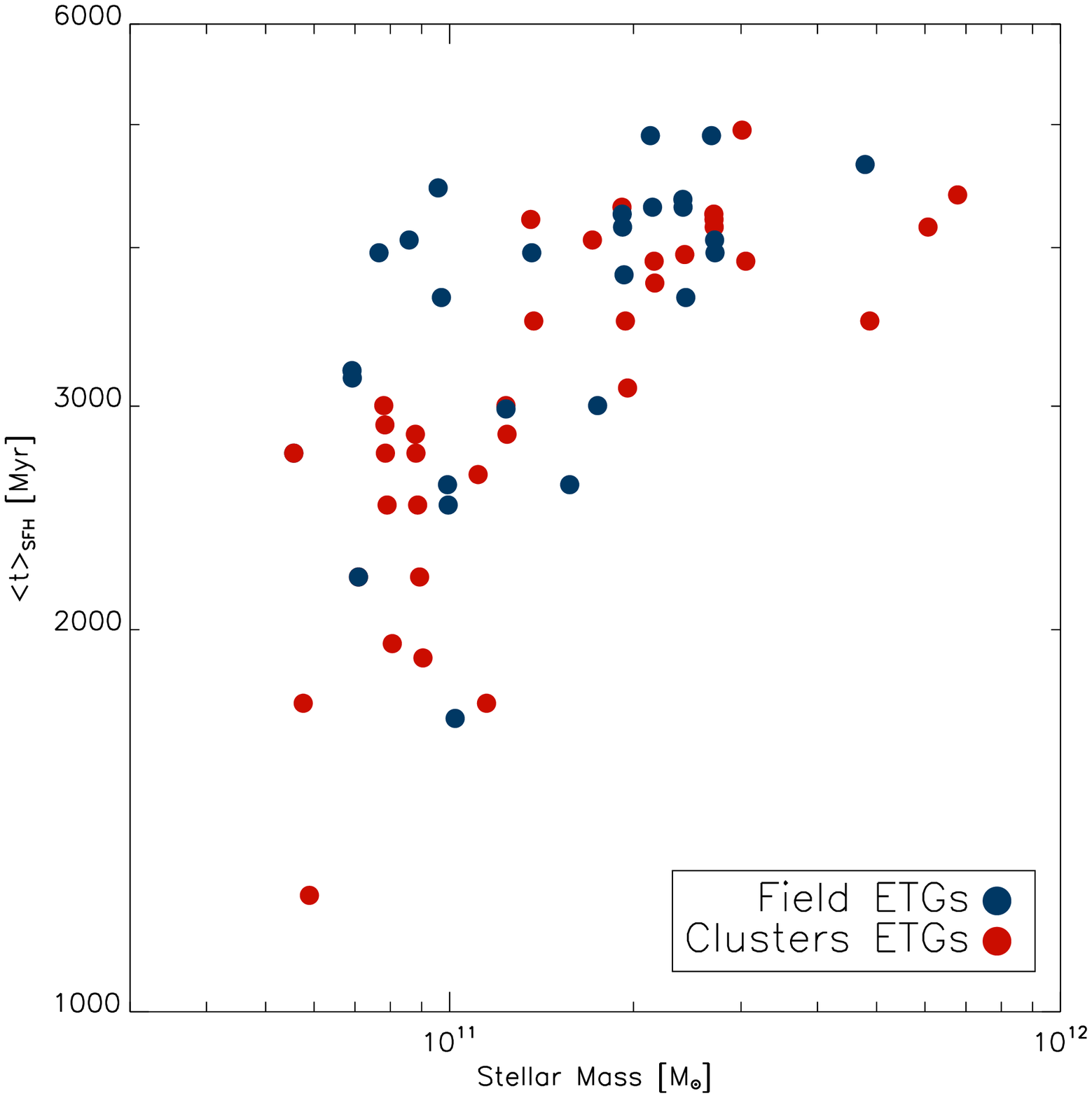}
\includegraphics[scale=.45,angle=90]{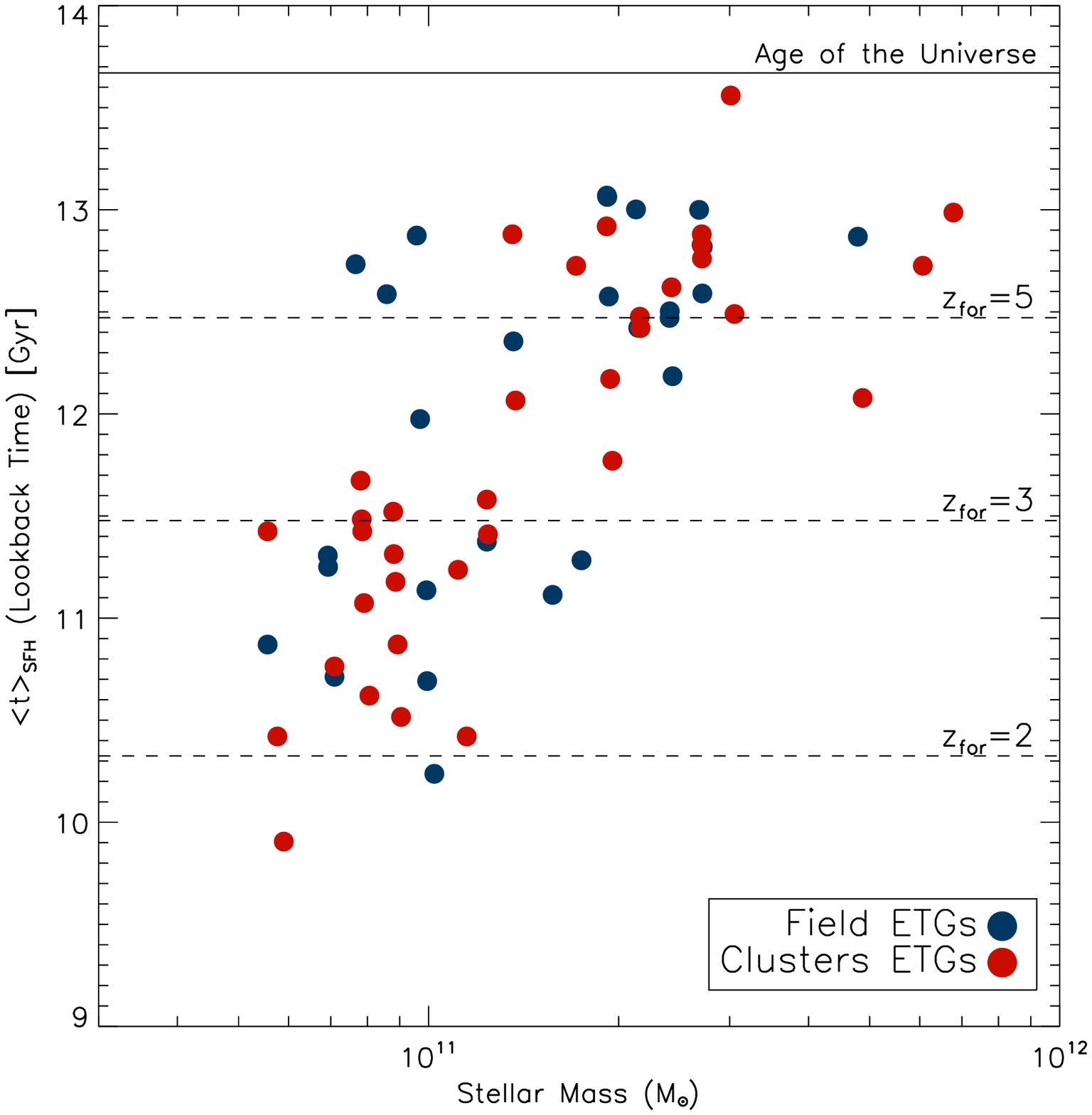}
\caption{{\it  Top  Panel:}  The  dependence  of  the  star  formation
  weighted ages on the galaxy  mass and environment.  No dependence of
  ages on the environment is found. {\it Bottom Panel:} The dependence
  of the lookback  time to star formation weighted  ages on the galaxy
  mass,  and environment. Uncertainties in  stellar masses and  ages are
  $\sim 0.15$ dex and $\sim 0.5$ Gyr.}
\label{ageclusfield}
\end{figure*}

\begin{figure*}
\epsscale{1.}
\includegraphics[scale=.35,angle=90]{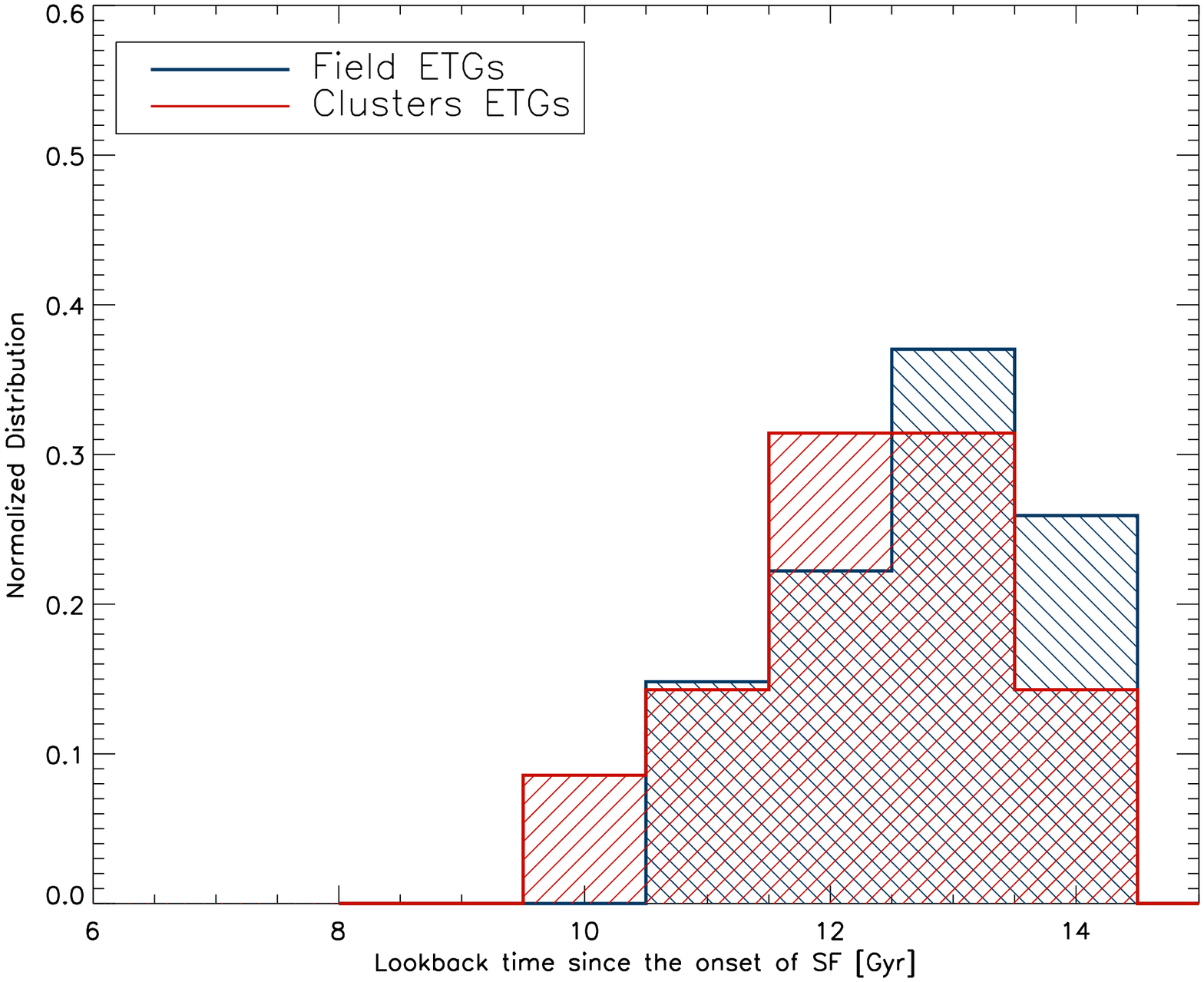}
\includegraphics[scale=.35,angle=90]{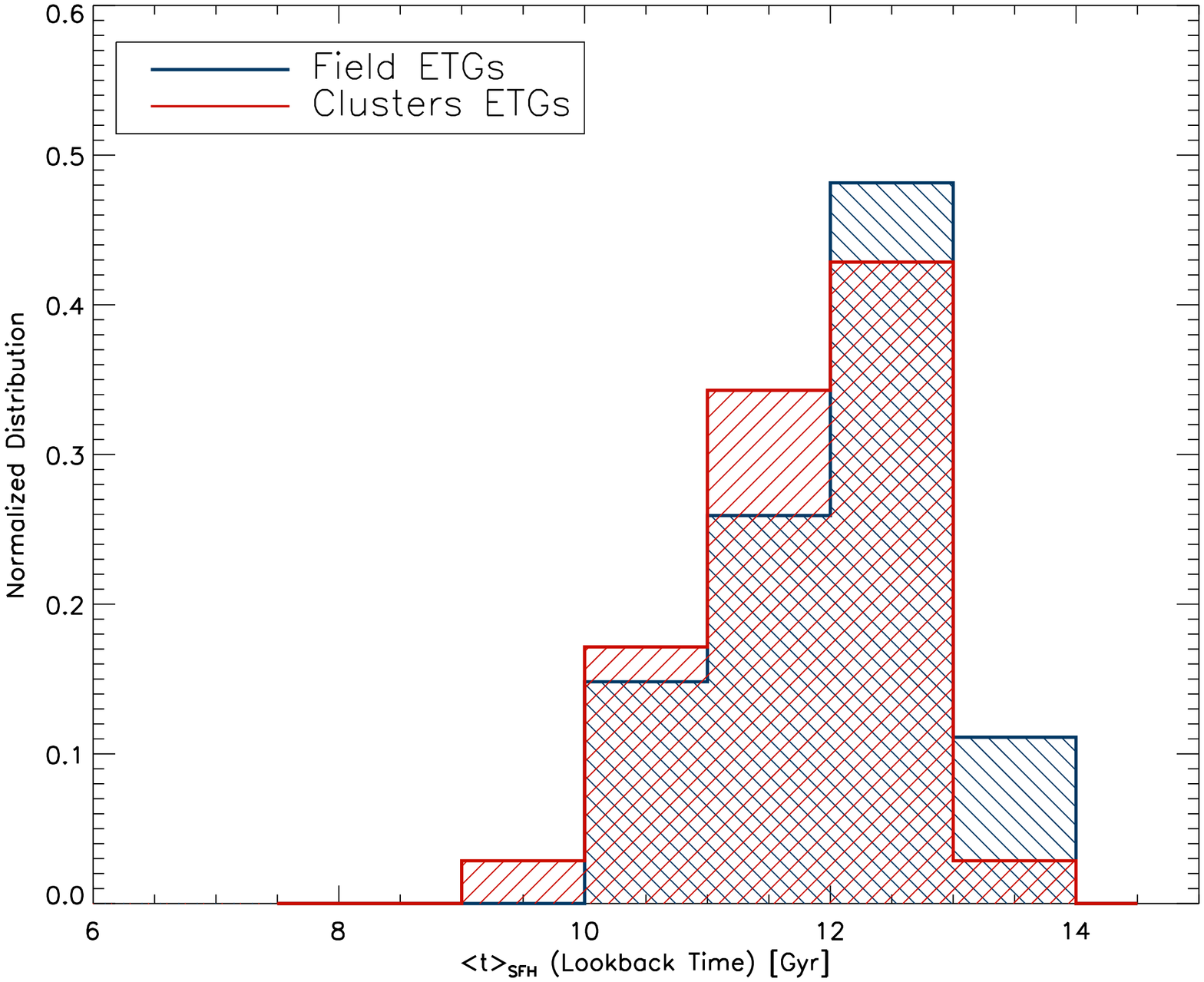}
\caption{Histograms of the field  (solid blue line) and cluster (solid
  red  line) lookback  times since  the onset  of star  formation (top
  panel) and of  the star formation weighted ages  (bottom panel).  No
  dependence of  the ETG ages with the  environment.  Uncertainties in
  age determination are $\sim 0.5$ Gyr.}
\label{histageclusfield}
\end{figure*}

Thus we  compare the distribution of the  star-formation weighted ages
($<t>_{SFH}$, top  panel of Fig.~\ref{ageclusfield}) as  a function of
stellar mass  in both environments.   We find the distributions  to be
overall  very similar,  as also  shown by  the histograms  of lookback
times   since   the   onset   of   star  formation   (top   panel   of
Fig~\ref{histageclusfield})  and  by the  histograms  of the  lookback
times  to $<t>_{SFH}$  (bottom  panel of  Fig~\ref{histageclusfield}).
Note that Mann-Whitney U-Test returns a median probability of 18\% and
the K-S statistics a 60\% of probability that the last two data values
are drawn  from the  same distribution.  These  results imply  that no
significant  delay  in  relative  age  is found  for  ETGs  in  either
environments.  Using  BC03 models,  we find that  $\sim 80$\%  of ETGs
have SF-weighted ages  in the range $3.5 \pm 1.0$  Gyr in both cluster
and  field,  in  qualitative   agreement  with  previous  CMR  studies
\citep{Blakeslee03, Mei06a,Mei06b}.

In  the bottom  panel  of Fig.~\ref{ageclusfield}  we  show, for  both
samples, the dependence  of star-formation weighted ages, $<t>_{SFH}$,
on stellar mass.  It is evident that the ages of ETGs only
depends on their mass, i.e.  the halo mass in which they reside, which
is  in agreement  with  the so--called  {\it  downsizing} scenario  of
galaxy  formation \citep{Cowie96}.   We note  that this  scenario have
also  been  reconciled, with  the  introduction  of  various forms  of
so-called   feedback   mechanisms,  with   the   recent  versions   of
semi-analytic    models    based    on   $\Lambda    CDM$    cosmogony
\citep{Delucia06, Bower06, Menci08}.

To summarize,  we find that  cluster galaxy (star  formation weighted)
ages   have   the   same   relative  distribution   as   their   field
contemporaries:  no significant  delay  in their  formation epochs  is
found within the  errors ($\sim 0.5$ Gyr). We  remind that this result
is   in  remarkably   good  agreement   with  the   ones   derived  by
\citet{vandokkum07} and \citet{dSA06} from  the evolution of the $M/L$
ratio. However our  method is able to extend this  kind of analysis to
less massive system than the typical  targets of FP studies at z $\sim
1$.

Despite of the fact that cluster and field galaxy formation epochs are
found to be similar, it could still be possible that the timescales of
their SFH  are significantly different.  As mentioned above,  the data
have shown that the distributions  of cluster and field optical colors
were slightly different.   As a function of the  stellar mass, cluster
galaxies are found to lie on a very tight red-sequence, while those in
the  field populate  the same  color-sequence with  a  slightly larger
scatter (Fig.~\ref{UMB_mass}).

\begin{figure*}
\epsscale{1.0}
\includegraphics[scale=.45]{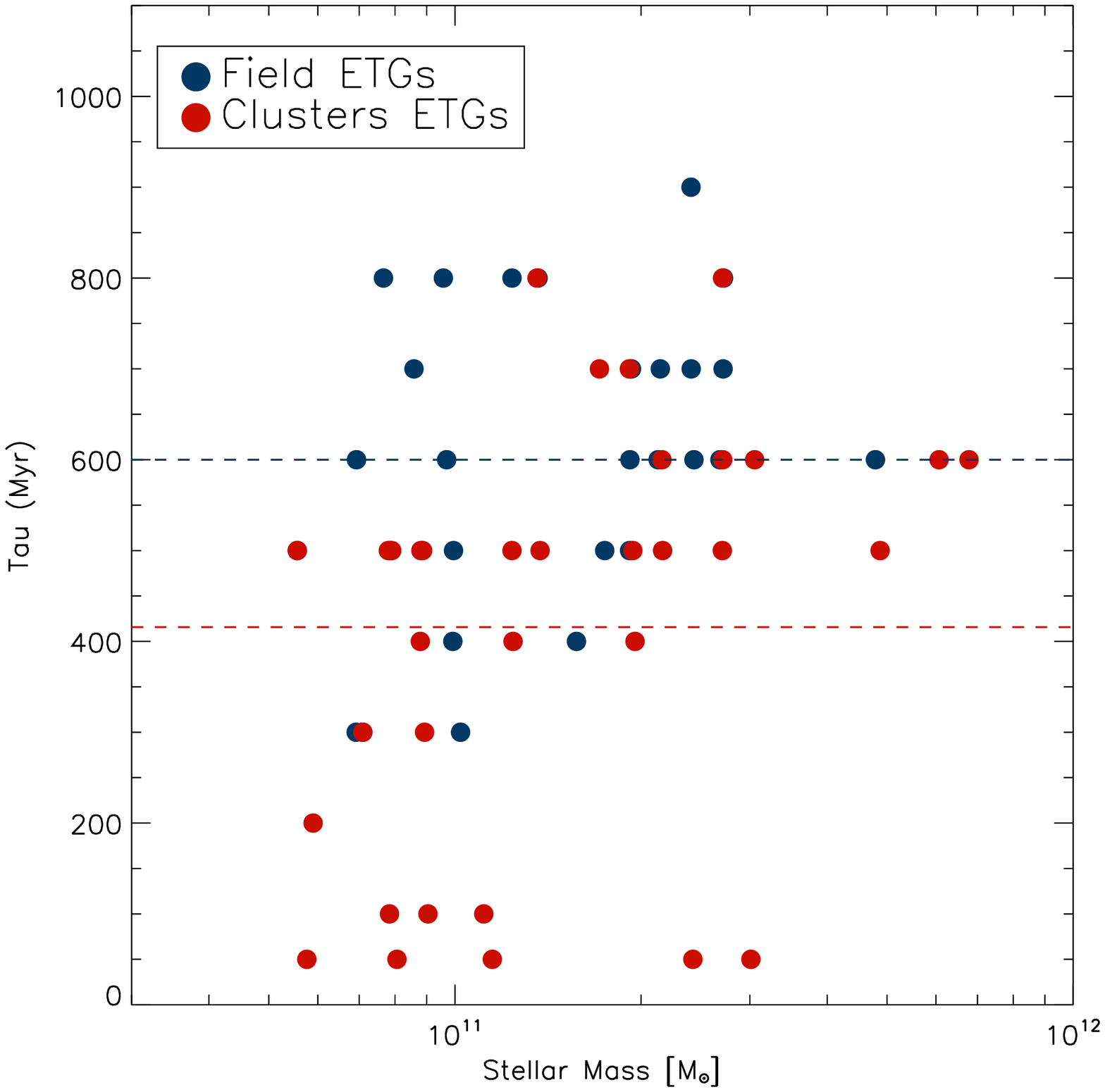}
\includegraphics[scale=.45,angle=90]{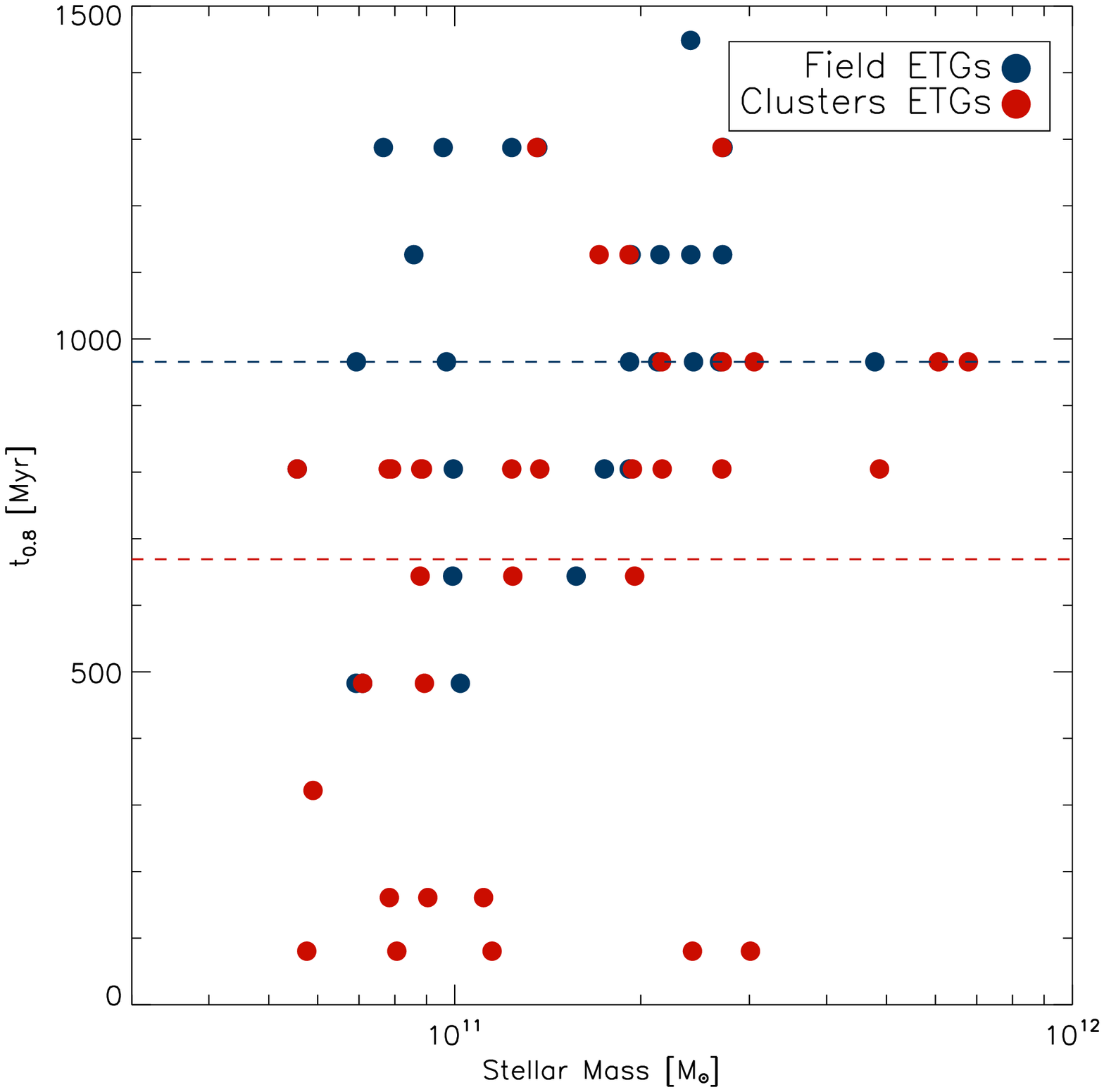}
\includegraphics[scale=.45,angle=90]{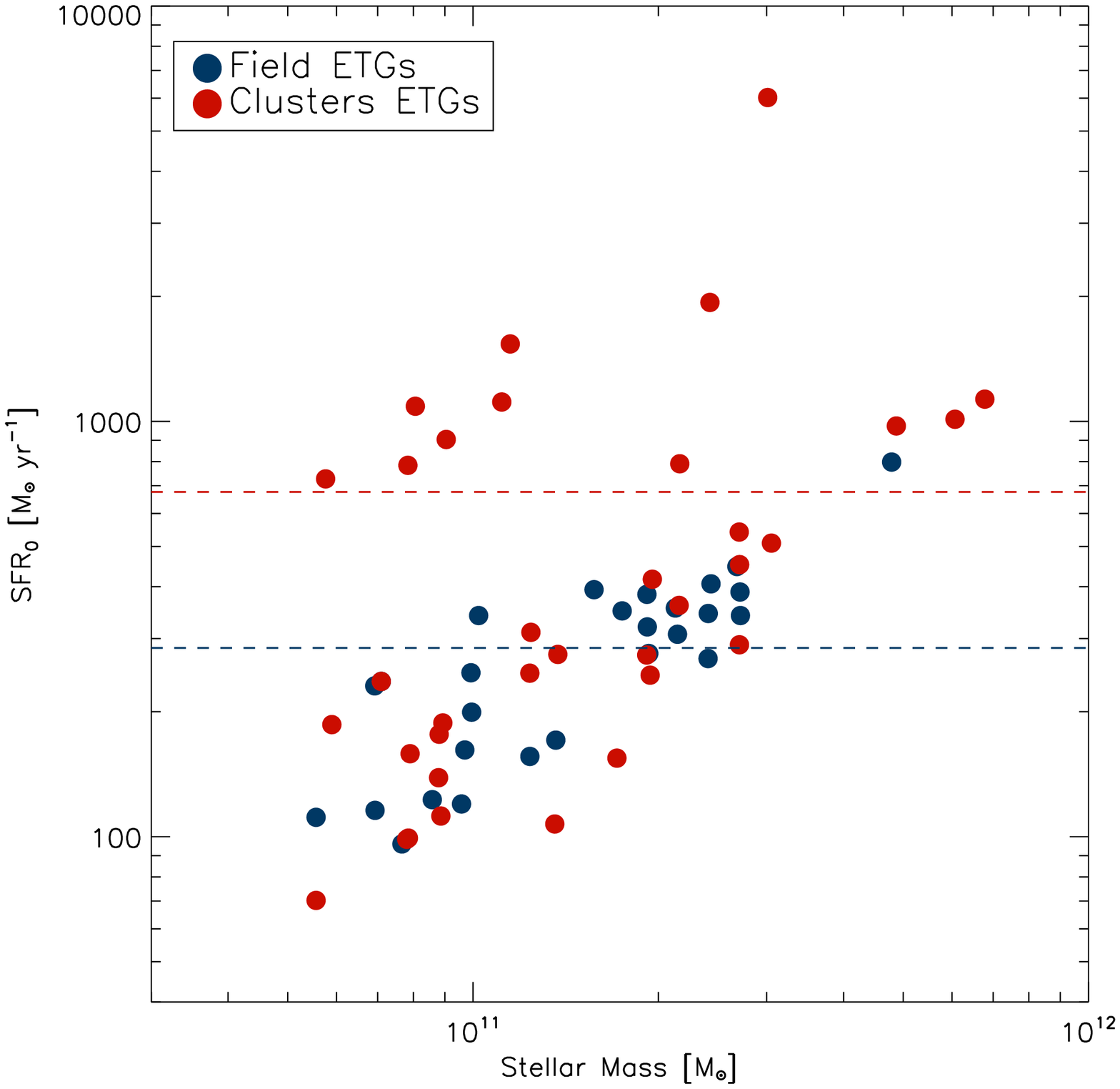}
\includegraphics[scale=.45]{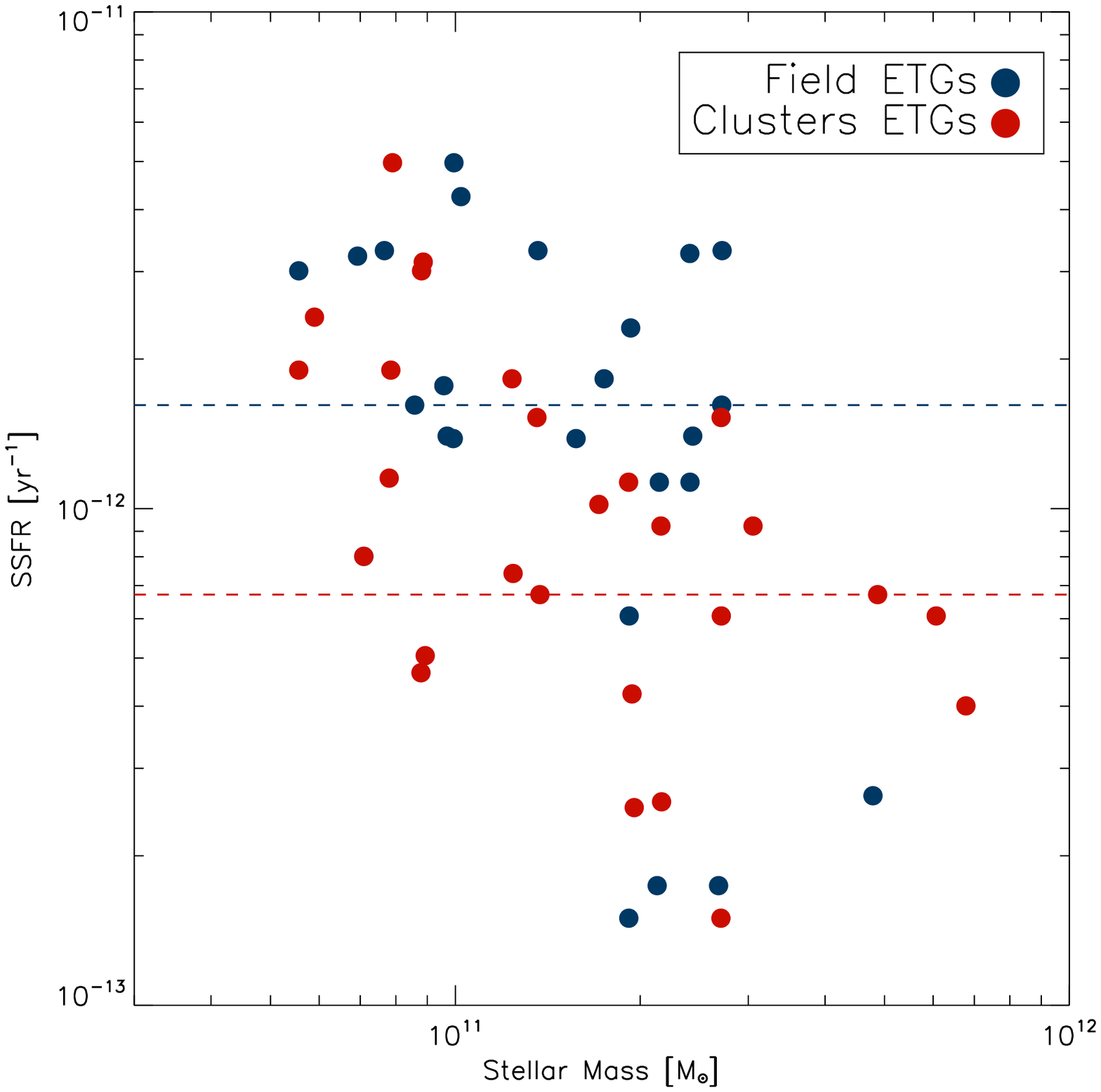}
\caption{{\it Top-left Panel:}  Formation {\it timescales}, $\tau$, of
  ETG  as a  function of  their  stellar mass  and environment.   {\it
    Top-right Panel:}  As a function of stellar  mass and environment,
  the diagram  of the  time, $t_{0.8}$, needed  for a given  galaxy to
  form  80\% of the  stars it  will have  at $z=0$.   {\it Bottom-left
    Panel:} Dependence of the  initial star formation rate, $SFR_{0}$,
  of ETGs  as a function of  their stellar mass  and environment. {\it
    Bottom-right Panel:} Specific SFRs,  $SSFR$, of ETGs as a function
  of their  stellar mass and  environment.  The mean error  in stellar
  age  is $0.5$  Gyr.   Uncertainties in  stellar  masses, $\tau$  and
  $t_{0.8}$,  are $\sim  0.15$ dex,  $\sim  0.2$ and  $\sim 0.4$  Gyr,
  respectively.  In  all panels,  as a function  of stellar  mass, the
  median values of  the different cluster (red dashed  line) and field
  (blue dashed  line) stellar population  parameter distributions are
  also indicated.}
\label{masstauall}
\end{figure*}

This piece of evidence finds  a natural explanation in the framework of
our    modeling.     As   shown    in    the    top-left   panel    of
Fig.~\ref{masstauall},  as a function  of stellar  mass, we  find that
field   ETGs   show  longer   SFH   timescales   than  their   cluster
contemporaries, which  tend to assemble  their mass with  the shortest
$\tau$ at  any given  mass. In fact  the Mann-Whitney  U-Test confirms
that  the  distributions  are  very  likely to  be  different  (median
probability of 0.07\%).   The Kolmogorov-Smirnov statistic also yields
an  associated probability that  the two  distribution of  $\tau$s are
drawn from the same distribution of just 0.8\%.

\begin{figure*}
\epsscale{1.}
\includegraphics[scale=.35,angle=90]{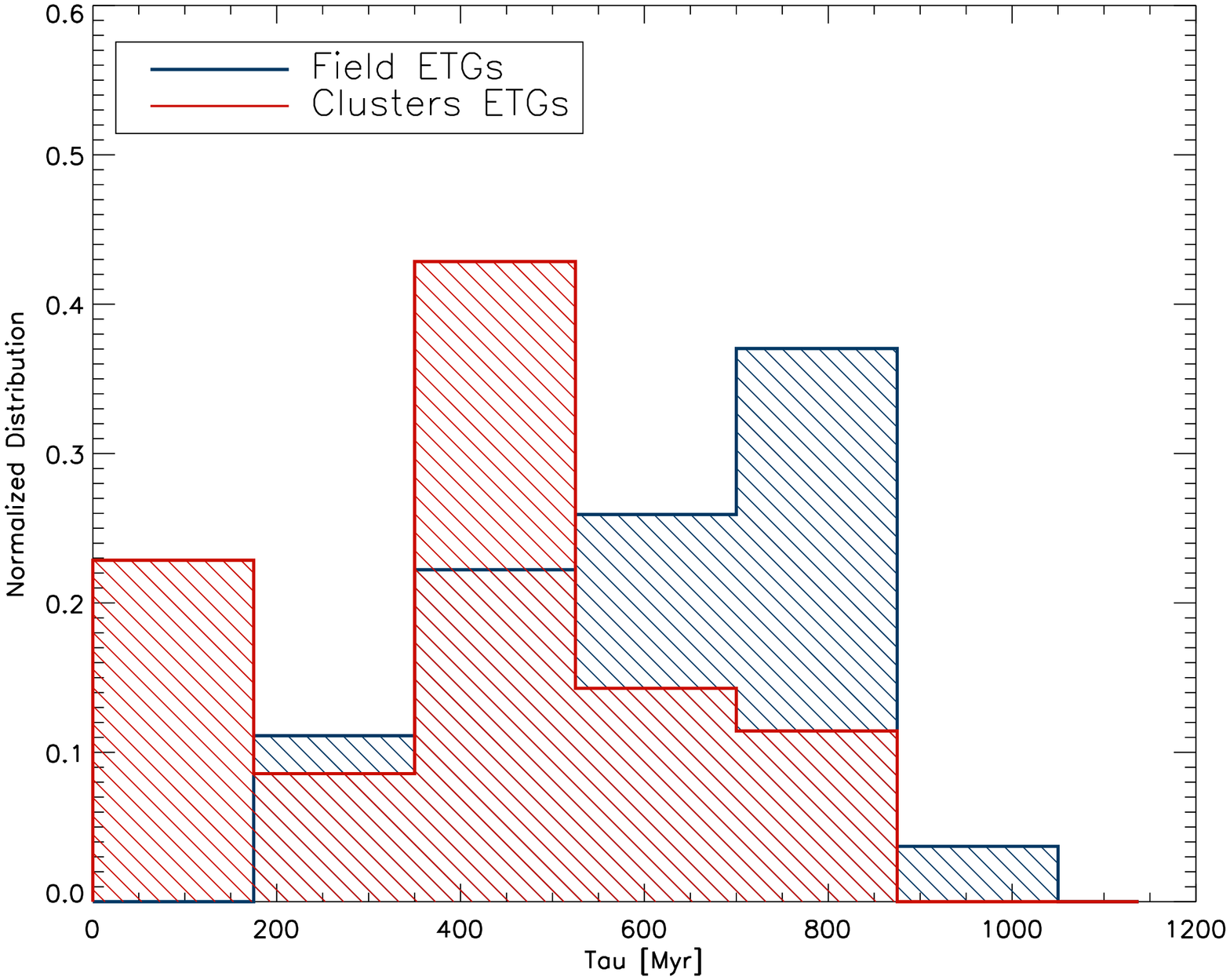}
\includegraphics[scale=.35,angle=90]{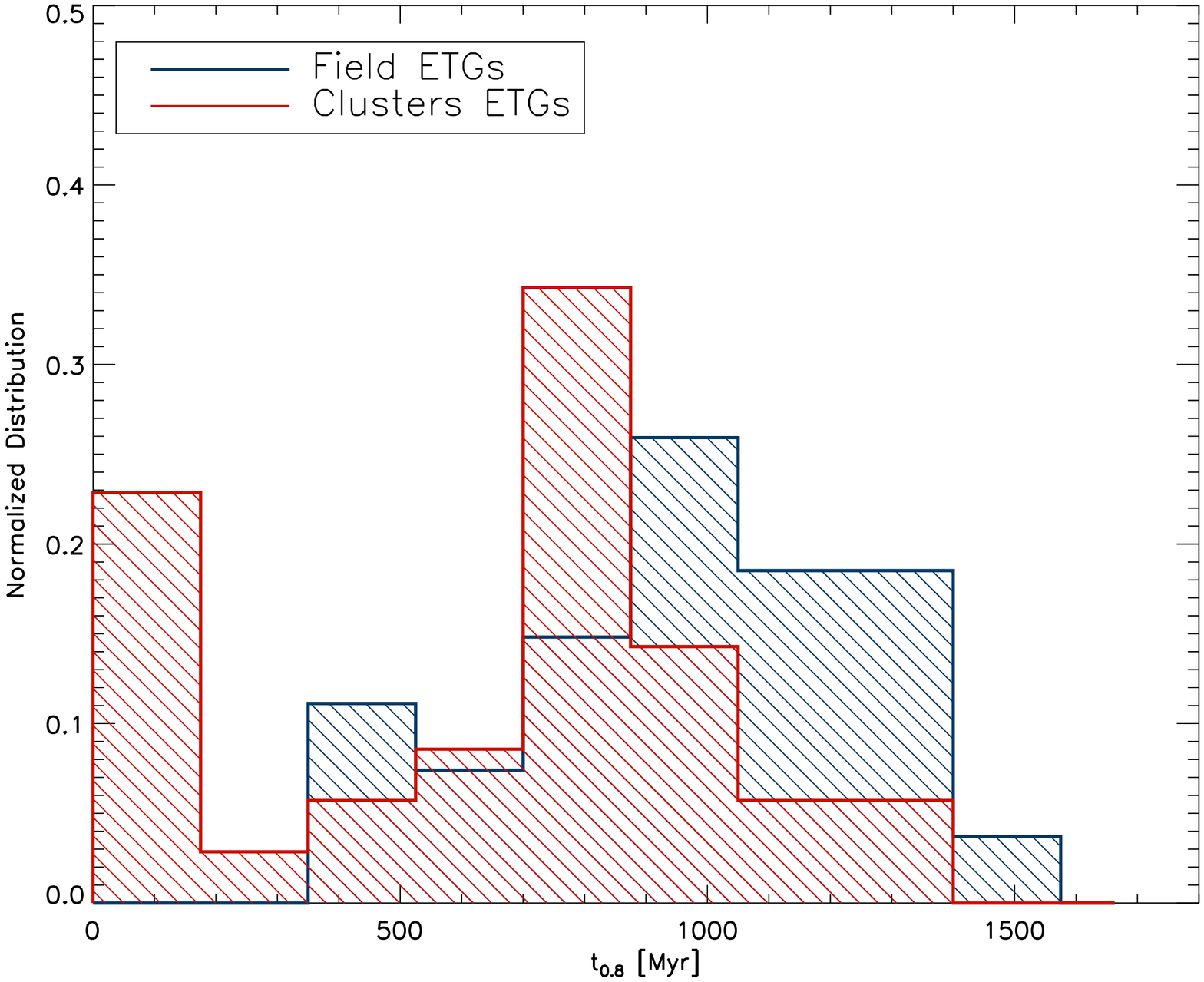}
\caption{{\it Top Panel:} Histogram of the formation {\it timescales},
  $\tau$, of ETGs in  both environments. {\it Bottom Panel:} Histogram
  of the times,  $t_{0.8}$, needed for a given galaxy  to form 80\% of
  the  stars   it  would  have   at  $z=0$,  for   both  environments.
  Uncertainties in the determination of $\tau$ and $t_{0.8}$ are $\sim
  0.2$ and $\sim 0.4$ Gyr, respectively.}
\label{histtauclusfield}
\end{figure*}

If galaxies continue to form  stars following the exponential decay of
Eq.~\ref{eq:sfhs}, we  can derive the  time, $t_{0.8}$ needed  to form
80\% of  the stars they  would have at  $z=0$.  As shown  by top-right
panel    of     Fig.~\ref{masstauall}    and    bottom     panel    of
Fig.~\ref{histtauclusfield}),  we find  that  $\sim 1$  Gyr after  the
onset of  star formation 75\%  of cluster ETGs have  already assembled
most of their  of their final mass.  By the same  time, less than 35\%
of field ETGs have.

The  bottom-left  panel of  Fig.~\ref{masstauall}  shows the  relative
distributions of inferred initial star formation rate, $SFR_{0}$,
as a function of stellar mass. More than 30\% of cluster galaxies have
higher   $SFR_{0}$  and   smaller   $\tau$s  than   any  other   field
contemporaries, indicating  they have experienced a  much more intense
star formation  at early times than  field galaxies.  As  a result, at
the time  of observation,  cluster ETG specific  star-formation rates,
defined as  the SFR  divided by the  galaxy stellar mass,  $SSFR$, are
lower  than  those  of  the  field ones  (bottom-right  panel  of
Fig.~\ref{masstauall}).  Indeed the  Mann-Whitney U-Test confirms that
the  distributions  of  $SSFR$  are  likely to  be  different  (median
probability  of 1.3\%).   The Kolmogorov-Smirnov  statistic  yields an
associated probability  of 0.5\% that the two  distribution of $\tau$s
are drawn from the same parent distribution.

As  summarized by the  histograms in  Fig.~\ref{histtauclusfield}, the
distributions of $\tau$ and $t_{0.8}$ in clusters and in the field are
found  to  be  very  different.   These  results  clearly  indicate  a
dependence  of ETGs star  formation histories  on the  environment, in
agreement with  previous studies based on a comparison of a single cluster 
to the field \citep{Gobat08, Rettura10, Menci08}.

As discussed in \citet{Rettura10}, we  recall that our field sample is
more deficient in  lower mass objects than the  cluster sample because
of a less extensive follow-up of spectroscopic targets in the field as
compared to that in the cluster.\\
However,  even if the  field sample  were corrected  for completeness,
this would  likely result in  a larger fraction  of field ETGs  at low
mass,  which  are  the ones  that  we  find  with longer  $\tau$s  and
$t_{0.8}$s, higher  $SSFR$s and lower $SFR_{0}$s.   Hence a correction
would  actually  amplify  the  difference  between  the  typical  mass
assembly  timescales  of  the  two  samples, and  so  not  affect  our
conclusions.

\section {Conclusions}
We  have studied  the  environmental dependence  of early-type  galaxy
stellar population  properties at $z \sim 1.3$  based on high-quality,
multi-wavelength photometry  available in  8-10 passbands from  $U$ to
$[4.5\mu m]$: sampling  the entire relevant domain of  emission of the
different stellar populations,  from rest-frame far-ultraviolet to the
infrared.

We have similarly analyzed  mass-selected  samples of ETGs  belonging to
the  cores  of  three   massive  clusters  (RXJ0849+4452  at  z=1.261,
RXJ0848+4453  at  z=1.273, RDCS1252.9-2927  at  z=1.237) and  compared
their stellar population properties with those measured in a similarly
selected sample of field contemporaries drawn from the GOODS-S survey.

We have  derived stellar masses,  (star formation weighted) ages, and star  formation histories,
parameterized as timescales,  $\tau$, from exponentially declining CSP
model templates built with BC03 models  for samples of massive ($M > 5
\cdot 10^{10} M_{\odot}$), passive (no-emission line in their spectra)
ETGs  at z$\simeq  1.3$.  Apart  from a  lower level  of spectroscopic
completeness for the least massive field galaxies, that we find not to
affect  our  conclusions,  our  sample  has  the  advantage  of  being
photometrically  complete at our  mass limit  and having  galaxy types
(passive or star-forming members) assigned spectroscopically.

This  work  extends the  analyses  performed  on  a single-cluster  by
\citet{Rettura10, Gobat08} to two more
clusters at $z\sim 1.3$: doubling the size of the cluster ETG sample.\\
Hence, on the basis of the new data we are able to affirm that:

\begin{itemize}
\item{}  we find  no  significant difference  in  the derived  stellar
  population properties  of $z \sim  1.3$ ETGs belonging  to different
  X-ray     luminous     clusters     (RXJ0849+4452 \&  RXJ0848+4453,
  RDCS1252.9-2927).
\end{itemize}

Moreover the  comparative statistical analysis performed  in this work
on  the extended  dataset  corroborates the  results  obtained in  our
previous study \citep{Rettura10}, that we summarize as follows:

\begin{itemize}
\item{}  Field ETGs galaxies  are distributed  around the  cluster ETG
  red-sequence,  although   they  seem  to  show   a  larger  scatter,
  especially at lower masses. We  will need more statistics and better
  estimates  of uncertainties  to draw  a more  stringent conclusions,
  since  the  cluster  and  field  overall  scatters  could  still  be
  consistent within  the uncertainties.   However, we remind  that the
  small  scatter  in  cluster  environments remains  a  challenge  for
  semi-analytical galaxy evolution models \citep{Menci08}.

\item{} We  find no significant  delay in the star  formation weighted
  ages of  massive ETGs observed in  all clusters and in  the field at
  z$\simeq 1.3$.  \\ The age of  ETGs increase with galaxy mass in all
  environments,  which  is  in  agreement with  the  {\it  downsizing}
  scenario.  The site of active  star formation must have shifted from
  the most massive  to the less massive galaxies as  a function of the
  cosmic time. The formation epochs of ETGs only depends on their mass
  and  not  on  the environment  they  live  in.   This result  is  in
  remarkably good agreement with  those obtained from the evolution of
  the $M/L$ ratio (e.g.  \citet{vandokkum07, dSA06}).

\item{}  However,  the  data  show  that cluster  and  field  SFHs  are
  significantly different. Field ETGs best-fit models span a different
  range  of timescales  than their  cluster contemporaries,  which are
  formed with the  shortest $\tau$ at any given  mass.  This result is
  quantitatively  consistent with  the predictions  of  current galaxy
  formation  models based  on  the latest  rendition of  semi-analytic
  models \citep{Menci08}.

\item{} We find  that 1 Gyrs after the onset  of star formation 75\%
  of cluster ETGs have already assembled 80\% (or more) of their final
  mass,  while,  by  the same  time,  less  than  35\% of  field  ETGs
  have. 

\item{} Accordingly, cluster ETGs specific star-formation rates at the
  time of observation are found to  be smaller than those of the field
  ones, implying  that the  last episode of  star formation  must have
  happened more  recently in  the field than  in the cluster.

\end{itemize}

While cluster  and field  galaxy observed at  z$\simeq 1.3$ form  at a
similar  epoch in  a  statistical sense,  a  high density  environment
appear to  be able to  trigger a much  more rapid and  homogenous mass
assembly  event  for  the  ETGs,  limiting the  range  of  possible
star-formation processes. In fact,  more than 30\% of cluster galaxies
are found  with higher initial  $SFR_{0}$ and smaller $\tau$s  than any
other field contemporaries, indicating they have experienced much more
intense star  formation at  early times than  field galaxies.   In low
density environments, this effect must  rapidly fade as ETGs display a
much broader range of possible star formation histories.

\acknowledgments  A.R.  is  grateful  to Gabriella  de Lucia,  Raphael
Gobat,  Roderik Overzier, Maurilio  Pannella, Veronica  Strazzullo and
Loredana Vetere  for useful discussions.

ACS was developed under NASA  contract NAS 5-32865.  This research has
been  supported  by the  NASA  HST  grant  GO-10574.01-A, and  Spitzer
program  20694.   The  {\it  Space  Telescope  Science  Institute}  is
operated by {\it AURA Inc.},  under NASA contract NAS5-26555.  Some of
the  data  presented herein  were  obtained  at  the {\it  W.M.   Keck
  Observatory}, which  is operated  as a scientific  partnership among
the {\it  California Institute of Technology}, the  {\it University of
  California}   and   the   {\it   National  Aeronautics   and   Space
  Administration}.  The Observatory was  made possible by the generous
financial  support of the  {\it W.M.   Keck Foundation}.   The authors
wish to  recognize and acknowledge the very  significant cultural role
and reverence that  the summit of Mauna Kea has  always had within the
indigenous  Hawaiian community.   We are  most fortunate  to  have the
opportunity to conduct observations from this mountain. Some data were
based on observations obtained  at the {\it Gemini Observatory}, which
is operated  by the  {\it AURA, Inc.},  under a  cooperative agreement
with the {\it NSF} on behalf of the {\it Gemini} partnership: the {\it
  National Science  Foundation} (United States), the  {\it Science and
  Technology Facilities  Council} (United Kingdom),  the {\it National
  Research  Council}   (Canada),  {\it  CONICYT}   (Chile),  the  {\it
  Australian Research Council} (Australia), {\it Ministerio da Ciencia
  e Tecnologia} (Brazil) and  {\it Ministerio de Ciencia, Tecnologia e
  Innovacion Productiva}  (Argentina),  Gemini  Science  Program  ID:
GN-2006A-Q-78.

\email{aastex-help@aas.org}.






\clearpage

\begin{table*}
  \caption{Cluster and Field photometric datasets.} 
\label{table:1} 
\centering
\begin{tabular}{l l l l l}     
\\
\hline\hline
 & Lynx E & Lynx W & CL1252 & CDFS \\
Filter & (Tel./Instr.) & (Tel./Instr.) & (Tel./Instr.) & (Tel./Instr.) \\
\hline\hline
u' & $Keck/$LRIS & $Keck/$LRIS & $VLT$/VIMOS (U) & $VLT$/VIMOS (U) \\
B  & -           & -           & $VLT$/FORS2     & $HST/ACS$/F435W  \\
V  & -           & -           & $VLT$/FORS2     & $HST/ACS$/F606W  \\
R  & $Keck/$LRIS & $Keck/$LRIS & $VLT$/FORS2     & -            \\
$i_{F775W}$ & $HST/ACS$ & $HST/ACS$ & $HST/ACS$ & $HST/ACS$ \\
$z_{F850LP}$ & $HST/ACS$ & $HST/ACS$ & $HST/ACS$ & $HST/ACS$ \\
J  & $KPNO$/FLAMINGOS  & $KPNO$/FLAMINGOS  & $VLT$/ISAAC ($J_{s}$)  & $VLT$/ISAAC \\
Ks & $KPNO$/FLAMINGOS  & $KPNO$/FLAMINGOS  & $VLT$/ISAAC & $VLT$/ISAAC \\
$[3.6\mu m]$ &  $Spitzer/IRAC$  &  $Spitzer/IRAC$  &  $Spitzer/IRAC$  &  $Spitzer/IRAC$\\
$[4.5\mu m]$ &  $Spitzer/IRAC$  &  $Spitzer/IRAC$  &  $Spitzer/IRAC$  &  $Spitzer/IRAC$\\
\hline
\end{tabular}
\end{table*}

\clearpage

\clearpage

\clearpage

\clearpage

\clearpage

\clearpage

\clearpage

\end{document}